\documentclass[aps,pra,twocolumn,showpacs]{revtex4}
\usepackage[pdftex]{graphicx}
\usepackage{amssymb}
\usepackage{amsmath}
\usepackage{bm}

\DeclareGraphicsExtensions{.pdf}

\begin{document}
\title{Thermal friction on quantum vortices in a Bose-Einstein condensate} 

\author{Geol Moon}
\thanks{G. Moon and W. J. Kwon contributed equally to this work.} 
\author{Woo Jin Kwon}
\thanks{G. Moon and W. J. Kwon contributed equally to this work.}
\author{Hyunjik Lee}
\author{Yong-il Shin}

\email{yishin@snu.ac.kr}

\affiliation{Department of Physics and Astronomy and Institute of Applied Physics, Seoul National University, Seoul 08826, Korea}

\begin{abstract}
We investigate the dissipative dynamics of a corotating vortex pair in a highly oblate axisymmetric Bose-Einstein condensate trapped in a harmonic potential. The initial vortex state is prepared by creating a doubly charged vortex at the center of the condensate and letting it dissociate into two singly charged vortices. The separation of the vortex pair gradually increases over time and its increasing rate becomes higher with increasing the sample temperature $T$. The evolution of the vortex state is well described with a dissipative point vortex model including longitudinal friction on the vortex motion. For condensates of sodium atoms having a chemical potential of $\mu\approx k_B\times 120$~nK, we find that the dimensionless friction coefficient $\alpha$ increases from 0.01 to 0.03 over the temperature range of 200~nK $<T<$ 450~nK.
\end{abstract}


\maketitle


At finite temperature a superfluid coexists with thermal excitations of the system, and dissipation can occur in the superfluid dynamics via interactions with the thermal component. In the framework of the two-fluid model, the thermal dissipation  in a nonequilibrium state was conceptualized as mutual friction between an inviscid superfluid and a viscous normal fluid, arising from the relative velocity of the two fluids~\cite{Hall56, Barenghi83,Schwarz85}. This concept has been successfully applied in  superfluid helium research~\cite{Donnelly,BDVreview}, in particular, enabling one to phenomenologically describe the dissipative motion of quantized vortices in quantum turbulence~\cite{Hanninen14,Berloff14}. However, microscopic and quantitative understandings of the mutual friction on the vortex motion have not yet been completely established.

Atomic Bose-Einstein condensates (BECs), being a theoretically tractable superfluid system, provide an ideal setting for the microscopic study of the dissipative vortex dynamics~\cite{Kobayashi06,Berloff07,White14}. In previous BEC experiments, the thermal damping of vortex states was investigated with rotating or turbulent condensates~\cite{Rosenbusch02,Abo-Shaeer02,Neely13,Kwon14} but the direct comparison between the experimental results and theories was limited due to the complexity and uncertainty of the prepared vortex states. For a quantitative study of the dissipative vortex dynamics, it is desirable to probe long-time dynamics of a well-defined vortex state although it is experimentally challenging due to the finite lifetime of samples as well as technical imperfections of the initial-state preparation. It is noted that even for a simple case of an oblate axisymmetric BEC containing a single quantum vortex, the damping rate of the vortex state and its temperature dependence were predicted differently in various theoretical approaches~\cite{Fedichev99,Schmidt03,Duine04,Madarassy08,Jackson09, Wright10,Rooney10,Allen13,Gautam14,Yan14}. The decay of a vortex state can provide a clean testbed for nonequilibrium, finite temperature theories for BECs~\cite{Rooney10,Allen13}.

In this paper, we investigate the long-time dynamics of a corotating vortex pair in a highly oblate Bose-Einstein condensate trapped in a harmonic potential. The corotating vortex pair is generated from the dissociation of a doubly charged vortex that is created at the center of a condensate with a topological phase imprinting method~\cite{Isoshima00,Leanhardt02,Shin04,Choi12}. When the two vortices undergo pair orbiting motion, the dissipation effect is unambiguously revealed with the increasing of the intervortex distance of the vortex pair. The pair separating rate is enhanced at higher temperatures, demonstrating the thermal nature of the dissipation. We show that the evolution of the vortex state is well described with a dissipative point vortex model including longitudinal friction on the vortex motion and we determine the friction coefficient from the temporal evolution of the pair separation for various temperatures. This work provides clear, quantitative information on the thermal dissipation in the vortex dynamics. Furthermore, it demonstrates the applicability of the concept of mutual friction to the atomic BEC system.

\begin{figure}
\includegraphics[width=7.5cm]{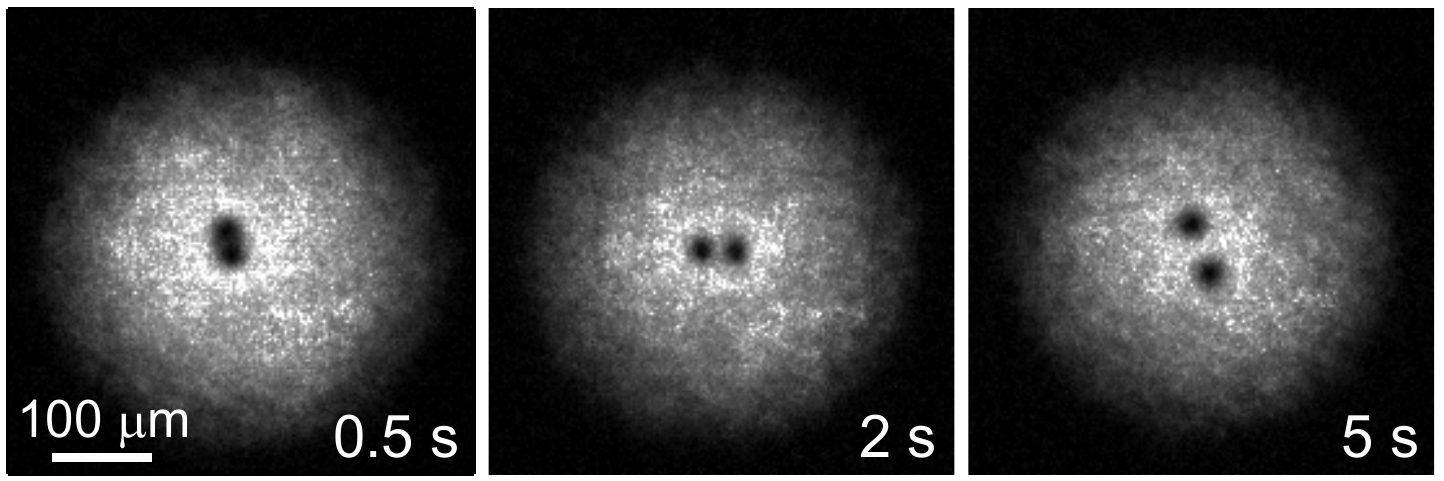}
\caption{Corotating vortex pair in a Bose-Einstein condensate. Images of the condensate at various hold times after creating a doubly charged vortex. The doubly charged vortex splits into two singly charged vortices, and at long times, the vortex pair becomes further separated.}
\end{figure}

\begin{figure*}
\includegraphics[width=14.5cm]{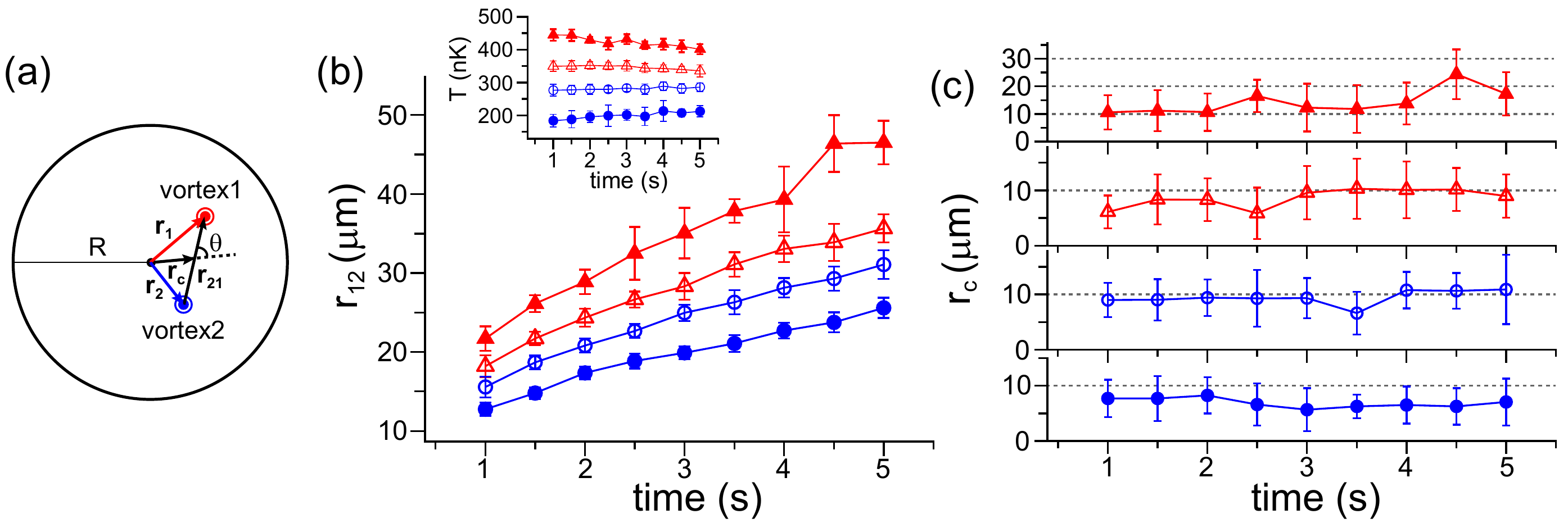}
\caption{(Color online) Evolution of the corotating vortex pair in a trapped Bose-Einstein condensate. (a) The vortex state is characterized by two configuration parameters: the separation distance $r_{12}$ of the two vortices and the radial position $r_c$ of the vortex pair center. The radius of the trapped condensate is denoted by $R$. Temporal evolutions of (b) $r_{12}$ and (c) $r_c$ for various sample temperatures. Each data point was obtained by ten measurements of the same experiment and its error bar indicates the standard deviation of the measurements. The inset in (b) shows the variation of the sample temperature $T$ during the evolution. In the measurements, $R\approx 76~\mu$m and the chemical potential of the condensate $\mu \approx k_B\times 120$~nK.}
\end{figure*}

The experimental apparatus was described in our previous work~\cite{Choi12,Choi13}. We first generate a Bose-Einstein condensate of $^{23}$Na atoms in the $|F=1,m_{F}=-1\rangle$ state in a pancake-shaped optical dipole trap and we apply a magnetic quadruple field $\textbf{B}=B_q(x\hat{x}+y\hat{y}-2z\hat{z})/2+B_z\hat{z}$, where the symmetric axis of the magnetic field is aligned to that of the optical trap. Including the Zeeman energy and the gravitational energy, the external potential for atoms is described as
\begin{eqnarray}
V(x,y,z)&=&\frac{m}{2}(\omega_x^2x^2+\omega_y^2y^2+\omega_z^2z^2)\\
&+&\frac{\mu_BB_q}{2}\sqrt{\frac{x^2+y^2}{4}+(z-z_m)^2}+mgz,\nonumber
\end{eqnarray}
where $m$ is the atomic mass, $(\omega_x,\omega_y,\omega_z)=2\pi \times (5.6, 8.4, 690)$~Hz are the trapping frequencies of the optical potential, $\mu_B$ is the Bohr magneton, $z_m=B_z/B_q$ is the axial position of the zero-field point of the magnetic field, and $g$ is the gravitational acceleration. Initially, $B_q=7.6$~G/cm and $z_m=130~\mu$m.

Using a topological phase imprinting method, we create a doubly charged vortex state~\cite{Leanhardt02,Shin04,Choi12}. We linearly decrease the bias field $B_z$ for 100~ms to move the zero-field point to $z_m=-280~\mu$m. While the zero-field point penetrates through the condensate, the atomic spin of the condensate adiabatically follows the local magnetic field and the condensate acquires a $4\pi$ superfluid phase winding around the penetrating line of the zero-field point due to the geometric Berry phase~\cite{Isoshima00,Berry}. After the field ramp, we decrease the field gradient $B_q$ to 4.6~G/cm for 100~ms and simultaneously change $B_z$ to make $z_m=-105~\mu$m~\cite{footnote3}.

After the sample preparation,  the hybrid trap has an average transverse trapping frequency $\omega_{r}=\sqrt{\frac{\mu_B B_q}{8m |z_m|} + \frac{\omega_{x}^2+\omega_{y}^2}{2}}= 2\pi\times 19.7$~Hz with a slight anisotropy of $|\omega_x^2-\omega_y^2|/(2\omega_r^2)\approx 0.05$. The highly oblate geometry with  $\omega_z/\omega_r\approx 35$ strongly suppresses vortex line excitations~\cite{Rooney11} and the vortex dynamics in the condensate is effectively two dimensional (2D). The condensate contains $N_0\approx 3.4\times10^6$ atoms and its radial extent is measured to be $R\approx76~\mu$m. The healing length is $\xi=\hbar/\sqrt{2m\mu}\approx 0.3~\mu$m at the condensate center, where the chemical potential $\mu=m\omega_r^2 R^2/2\approx k_B\times 120$~nK. The sample temperature $T$ is determined from the condensate fraction $N_0/N$~\cite{Naraschewski98,Gerbier04}, where $N$ is the total atom number of the sample.

Vortices are detected by taking a time-of-flight absorption image of the condensate~\cite{footnote2}.  In the imaging, the radial extent of the condensate is expanded by a factor of 1.95 and the full width at half maximum (FWHM) of the density-depleted core of a singly charged vortex is measured to be about 10~$\mu$m in the center region of the condensate.

The doubly charged vortex splits into a pair of vortices~\cite{Shin04,Virtanen06} and the two vortices become spatially resolved at a hold time $t\approx 0.5$~s after the phase imprinting~(Fig.~1). At longer hold times, they evolve to be further separated, indicating that dissipation takes place in the vortex dynamics. The phase imprinting process causes breathing mode oscillations of the condensate because of the modulations of the transverse magnetic confinement. In this work, we restrict our study to the time period of $t>1$~s when the breathing mode excitations are sufficiently damped out~\cite{footnote1}.

The nondissipative dynamics of a corotating vortex pair in a harmonic potential was investigated in a recent experiment~\cite{Navarro13}, where it was shown that the vortex motion is well described by a point vortex model~\cite{Middelkamp10b,Torres11,Middelkamp11}. In a trapped condensate, the velocity of the $i$th vortex ($i,j=1,2$) is given by
\begin{eqnarray}\label{eq17}
\textbf{v}^0_{i}=\Omega(r_{i})(\hat{\textbf{z}}\times \textbf{r}_{i})+\frac{b}{2}\frac{\hbar}{m r_{ji}^2}(\hat{\textbf{z}}\times \textbf{r}_{ji}),
\end{eqnarray}
where $\mathbf{r}_{i}$ is the vortex position with respect to the condensate center and $\textbf{r}_{ji}=\textbf{r}_i-\textbf{r}_j$. The first term corresponds to the precession motion around the trap center due to the inhomogeneous density distribution of the condensate. For a pancake-shaped condensate, $\Omega(r)=\Omega_0/ (1-r^2/R^2)$ with $\Omega_0= \frac{3\hbar}{2m R^2}\ln(\frac{R}{\xi})$~\cite{Fetter01}. The second term is the superfluid flow field generated by the other vortex, which drives the vortex pair to orbit around each other. The parameter $b$ is the modification factor of the interaction strength between the vortices in comparison to that in the homogeneous condensate and $b=1.35$ for two vortices in a harmonic trap~\cite{Middelkamp10b}. In the symmetric case with $\textbf{r}_1=-\textbf{r}_2$, the vortex pair undergoes a circular orbiting motion whose angular frequency is $\omega_p=\Omega(\frac{r_{12}}{2})+b\hbar/(m r_{12}^2)$.

We investigate the dissipation effect in the dynamics of the corotating vortex pair by measuring the evolutions of the two configuration parameters: the intervortex distance $r_{12}=|\mathbf{r}_2-\mathbf{r}_1|$ and the radial position of the center of the vortex pair $r_c=|\mathbf{r}_2+\mathbf{r}_1|/2$~[Fig.~2(a)]. Assuming that the relative positions of the vortices are preserved in the condensate during the expansion in the imaging, we determine the \textit{in situ} values of $r_{12}$ and $r_c$ by dividing the values measured from the image by the expansion factor. Measurement results for various sample temperatures are displayed in Figs.~2(b) and 2(c). Note that in order to preserve the spatial size $R$ of the condensate, we varied the sample temperature with a different total atom number $N$ while maintaining the condensate atom number $N_0$, which was achieved by adjusting the initial thermal atom number and the evaporation cooling efficiency in the sample preparation sequence.

We observe that the increasing rate of the pair separation $r_{12}$ as well as its initial value at $t=1~$s becomes higher at higher temperatures. This clearly demonstrates the thermal nature of the dissipation in the vortex dynamics. When the two vortices orbit around each other, the energy of the vortex state can be also dynamically dissipated via sound emission from the accelerating vortices~\cite{Lundh00,Parker04,Barenghi05,Billam15}. In our case, $r_{12}>10~\mu$m and the wavelength of the generated sound wave is estimated to be $\lambda\sim 2\pi c_s /\omega_p \geq~1$~mm, where $c_s=\sqrt{\mu/m}$ is the speed of sound. Since $\lambda \gg R$, we expect that the dynamical dissipation effect is negligible~\cite{Fetter01}.

One remarkable observation in the measurement results is that fluctuations of the pair separation $r_{12}$ are noticeably small although the pair center position $r_c$ scatters in a relatively large area. This indicates that the splitting process of the doubly quantized vortex was quite deterministic and furthermore, the subsequent separating dynamics of the vortex pair is not significantly affected by the pair center position. This is the crucial feature of our experiment, which allows the precise characterization of the long-time vortex dynamics in spite of technical imperfections in the initial state preparation.


To elucidate the observed dissipative dynamics of the vortex pair, we adopt a dissipative point vortex model that includes the effect of longitudinal friction in the vortex motion~\cite{Madarassy08,Jackson09,Yan14,Billam15}. The friction force is presumed to be proportional to the relative velocity of the vortex to the thermal component of the system~\cite{Hall56,Schwarz85}. Here we assume a stationary thermal cloud. Our vortex creation method does not affect the thermal cloud and moreover, the slight anisotropy of the transverse trapping potential would help lock the thermal cloud to the trap~\cite{Wright10,Zhuravlev01}. This friction force needs to be balanced with a Magnus force that would arise from a change in the vortex motion and the resultant motion of the vortex is given by 
\begin{equation}
\mathbf{v}_i=\mathbf{v}^0_i-\alpha \hat{\mathbf{z}}\times\mathbf{v}^0_i,
\end{equation}
where $\alpha$ is the dimensionless coefficient that characterizes the amplitude of the friction. Figure 3 shows vortex trajectories obtained from numerical simulations of the model for our experimental parameters and with $\alpha=0.02$. Since the local density of the thermal cloud varies over the trapped sample, $\alpha$ should be position dependent. In our model study, we assume constant $\alpha$, which would be valid in the center region of the harmonically trapped sample, and neglect the slight trap anisotropy.

\begin{figure}
\includegraphics[width=8.4cm]{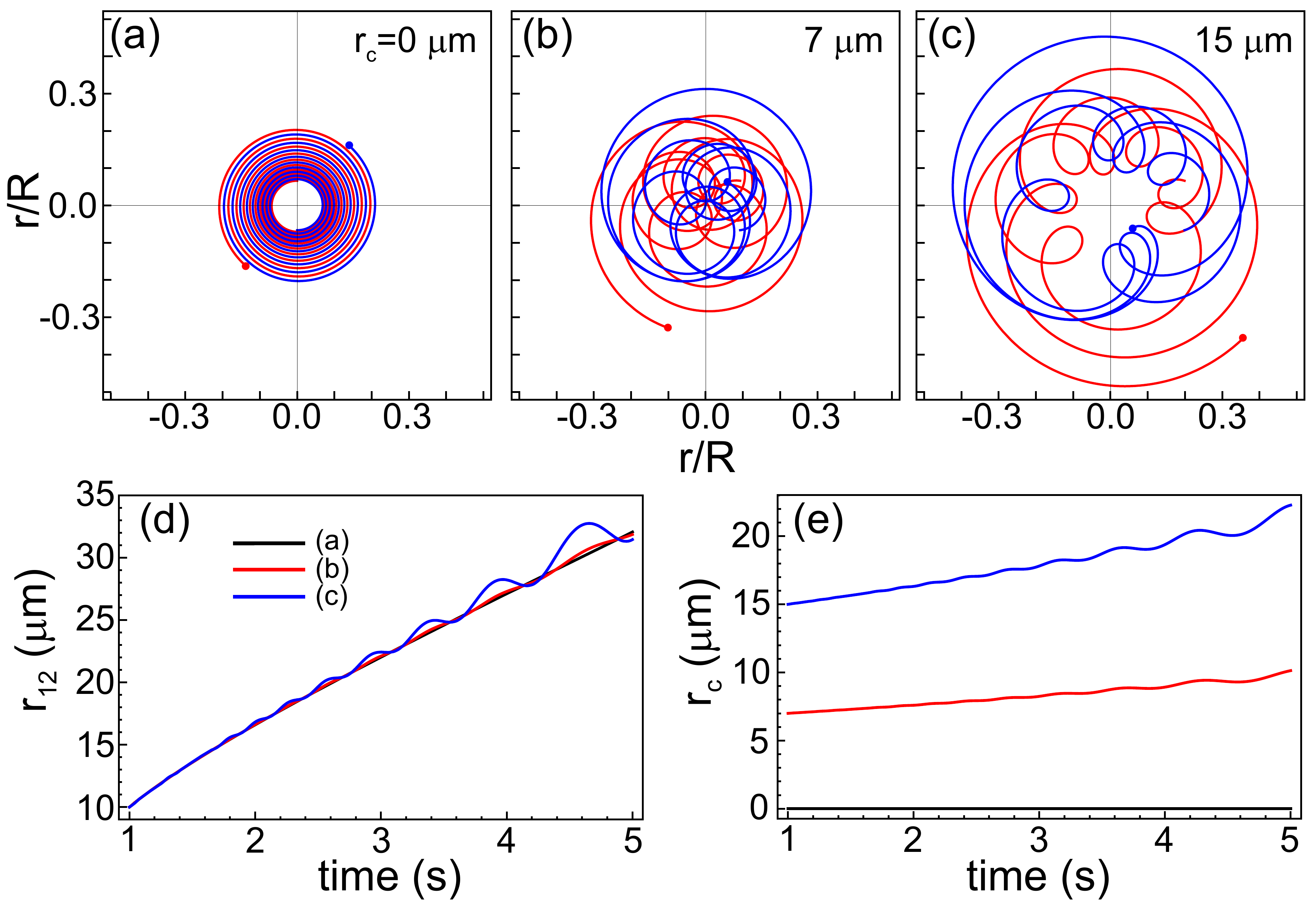}
\caption{(Color online) Numerical simulation of the dissipative point vortex model. (a)--(c) Trajectories of the two vortices for various initial conditions at $t=1$~s: $r_{12}=10~\mu$m, $\theta=\pi/2$, and $r_c=0~\mu$m in (a), 7~$\mu$m in (b), and 15~$\mu$m in (c). In the simulation, the friction coefficient $\alpha=0.02$, $R=76~\mu$m, and $\Omega_0/2\pi=0.64$~Hz. (d),(e) Corresponding evolutions of $r_{12}$ and $r_c$, respectively, for the vortex trajectories in (a)--(c).}
\label{Fig4}
\end{figure}

When $r_c=0$, the two vortices show symmetric spiralling-out trajectories, as expected~[Fig.~3(a)]. The growth behavior of $r_{12}$ resembles the experimental data in Fig.~2(b). In the asymmetric cases with $r_c\neq 0$~[Figs.~3(b) and 3(c)], the vortex trajectories appear complicated but the smooth evolutions of $r_{12}$ and $r_c$, shown in Figs.~3(d) and 3(e), respectively, suggest that they can be understood as relaxing epicyclic trajectories where a pair orbiting motion is superposed upon a slow precession of the pair center. In particular, we see that apart from the small oscillations, the evolutions of $r_{12}$ for $r_c\neq 0$ are almost identical to that in the symmetric case with $r_c=0$ [Fig.~3(d)].

The oscillations of $r_{12}$ and $r_c$ originate from the radial dependence of $\Omega(r)$, i.e., $\Omega(r_1) \neq \Omega(r_2)$. When the two vortices are located near the condensate center, i.e., $r_i\ll R$, the precession frequency can be approximated to be $\Omega(r)\approx \Omega_0$ and then the motions of $r_{12}$ and $r_c$ are effectively decoupled as 
\begin{eqnarray}
\frac{d\mathbf{r}_{12}}{dt}&=&\omega_{p0}\big[\alpha +\hat{\mathbf{z}}\times \big]\mathbf{r}_{12}, 
\\
\frac{d\mathbf{r}_{c}}{dt}&=&\Omega_0\big[\alpha+ \hat{\mathbf{z}}\times \big]\mathbf{r}_{c},
\end{eqnarray}
where $\omega_{p0}=\Omega_0+b\hbar/(m r_{12}^2)$.  This feature of the vortex motion seems to account for the experimental observation of small fluctuations in $r_{12}$ despite a relatively large scattering of $r_c$. Furthermore, Eq.~(4) suggests that in the center-region limit the friction coefficient $\alpha$ can be reliably determined solely from the evolution of the pair separation $r_{12}$.

\begin{figure}
\includegraphics[width=7.4cm]{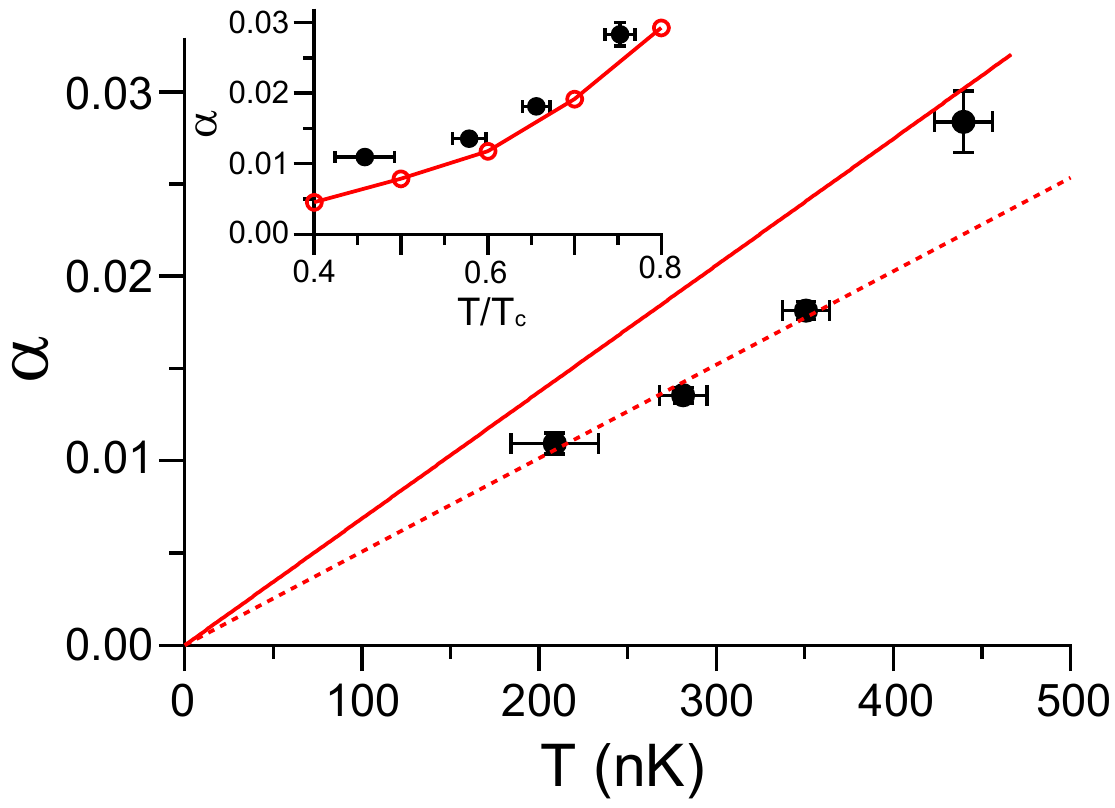}
\caption{(Color online) Friction coefficient $\alpha$ as a function of $T$.  Using Eq.~(4), $\alpha$ was determined from the experimental data in Fig.~2(b) (see the text for details). The solid line is the theoretical prediction of Fedichev \textit{et al.}~\cite{Fedichev99} and the dashed line is a linear fit to the three lowest-temperature data points. The inset displays $\alpha$ as a function of $T/T_c$. $T_c$ is the critical temperature of the trapped sample. The open circles show the numerical result of Jackson \textit{et al.}~\cite{Jackson09}, which was obtained for the decay of a single vortex in rubidium gases at fixed $T_c$.}
\end{figure}

Motivated by this finding, we determine the friction coefficient $\alpha$ by fitting the evolution curve of $r_{12}(t)$ obtained from Eq.~(4) to the experiment data in Fig.~2(b). Our analysis was restricted to the data in the range of $20~\mu$m$<r_{12}<33~\mu$m, where $r_c<20~\mu$m and so $r_i/R\leq 0.4$, marginally satisfying the center-region approximation. The measured value of $\alpha$ increases from 0.01 to 0.03 over the temperature range of 200~nK $<T<$ 450~nK (Fig.~4). The magnitude of $\alpha$ is comparable to the theoretical estimation of Fedichev \textit{et al.}~\cite{Fedichev99},  $\alpha \approx (n_\textrm{th}/n_s)\sqrt{\mu/(k_B T)}\sim~0.1\times \frac{4\pi a m^{1/2}}{\hbar \mu^{1/2}} k_B T$, where $n_\textrm{th}$ and $n_s$ are the densities of the thermal and superfluid components, respectively, and $a$ is the scattering length of the atoms. Intriguingly, when $\alpha$ is replotted as a function of $T/T_c$, where $T_c$ is the critical temperature of the trapped sample, we find it in good agreement with the numerical result of Jackson \textit{et al.}~\cite{Jackson09} (Fig.~4 inset). Its quantitative comparison should be restrictively made because the numerical result was obtained for the decay of a single vortex in condensates of $^{87}$Rb atoms at fixed $T_c$, i.e., having $\mu$ varying with $T$, whereas $\mu$ is constant in our measurements.

The temperature dependence of the mutual friction has been an important but unresolved issue in previous theoretical studies~\cite{Fedichev99,Duine04,Madarassy08,Kobayashi06,Berloff07,Jackson09}. Our result seems to reveal that the longitudinal friction coefficient $\alpha$ grows slightly faster than linearly with the temperature. However, it should be pointed out that in our analysis, the temperature dependence of the precession frequency $\Omega$ and the vortex interaction parameter $b$ is neglected. Indeed, it was theoretically predicted that $\Omega$ has a weak positive dependence on temperature~\cite{Isoshima04,Jackson09,Wild09,Allen13}. It is also important to recognize that the temperature dependence of $\Omega$ and $b$ is intricately related with the transverse friction which is not considered in our dissipative point vortex model~\cite{Kobayashi06,Berloff07,Madarassy08,Jackson09}. Further elaborated analysis including these effects as well as the realistic experimental condition such as the small trap anisotropy~\cite{Lundh00,Wright10,Zhuravlev01,stockhofe11} and the inhomogeneous density distribution of the thermal cloud~\cite{footnote4}  would solidify the quantitative interpretation of our measurement results. 

In conclusion, we have investigated the thermal dissipation in the dynamics of a corotating vortex pair in a BEC, in particular, in the view of the mutual friction on the vortex motion. The separation of the vortex pair provides a clear measure of the dissipation effect in the vortex dynamics, which allows us to determine the friction coefficient and its temperature dependence. We expect that the quantitative information on the thermal dissipation will give direct insight into the results of our previous experiment~\cite{Kwon14} where the relaxation of superfluid turbulence was investigated in terms of the decay rate of the vortex number of turbulent BECs.


We thank Sang Won Seo for experimental assistance. This work was supported by the National Research Foundation of Korea (Grant No. 2011-0017527).



\begin{references}
\bibitem{footnote3} This field adjustment is to weaken the downward magnetic force. Otherwise, the magnetic force, together with the gravitational force, would significantly reduce the trap depth, leading to undesired evaporation cooling of the sample.
\bibitem{footnote2} In releasing the trapping potential, we first turned off the magnetic field and then released the optical potential, which was helpful to improve the visibility of the vortex core~\cite{Kwon14}.
\bibitem{footnote1}  At $t=0.5$~s, the amplitude of the breathing mode oscillations was measured to be about 5\% of the condensate radius and its damping rate was about 2.5~s$^{-1}$.
\bibitem{footnote4} At $T=450$~nK ($k_B T/\mu\approx 3.8$), the relative thermal density $n_\textrm{th}/n_s$ is estimated to vary by 30\% over the center region of $r_i/R\leq 0.4$.
\end{references}
\end{document}